# Computational modeling of Human-nCoV protein-protein interaction network

**Sovan Saha**[1,+], **Anup Kumar Halder**[2,+], **Soumyendu Sekhar Bandyopadhyay**[2], **Piyali Chatterjee**[3], **Mita Nasipuri**[2], and **Subhadip Basu**[2,*]

[1]Dr. Sudhir Chandra Sur Degree Engineering College, Computer Science and Engineering, Kolkata, 700074, India
[2]Jadavpur University, Computer Science and Engineering, Kolkata, 700032, India
[3]Netaji Subhash Engineering College, Computer Science and Engineering, Kolkata, 700152, India
*subhadip.basu@jadavpuruniversity.in
+Equal Contribution, both shared 1st authorship

## ABSTRACT

COVID-19 has created a global pandemic with high morbidity and mortality in 2020. Novel coronavirus (nCoV), also known as Severe Acute Respiratory Syndrome Coronavirus 2 (SARS-CoV2), is responsible for this deadly disease. International Committee on Taxonomy of Viruses (ICTV) has declared that nCoV is highly genetically similar to SARS-CoV epidemic in 2003 (∼ 89% similarity). Limited number of clinically validated Human-nCoV protein interaction data is available in the literature. With this hypothesis, the present work focuses on developing a computational model for nCoV-Human protein interaction network, using the experimentally validated SARS-CoV-Human protein interactions. Initially, level-1 and level-2 human spreader proteins are identified in SARS-CoV-Human interaction network, using Susceptible-Infected-Susceptible (SIS) model. These proteins are considered as potential human targets for nCoV bait proteins. A gene-ontology based fuzzy affinity function has been used to construct the nCoV-Human protein interaction network at ∼ 99.98% specificity threshold. This also identifies the level-1 human spreaders for COVID-19 in human protein-interaction network. Level-2 human spreaders are subsequently identified using the SIS model. The derived host-pathogen interaction network is finally validated using 7 potential FDA listed drugs for COVID-19 with significant overlap between the known drug target proteins and the identified spreader proteins.

## Introduction

COVID-19 evolved in the Chinese city of Wuhan (Hubei province)[1]. The first case of human species getting affected by nCoV was observed on 31 December 2019[2]. Soon it expands its adverse effect on almost all nations within a very short span of time[3]. World Health Organization (WHO) observes that the massive disastrous outbreak of nCoV is mainly due to mass community spreading and declares a global health emergency on 30 January 2020[4]. After proper assessment, WHO presumes its fatality rate to be 4%[5] which urges the global researchers to work together to discover a proper treatment for this pandemic[6,7]. Coronaviridae is the family to which a corona virus belongs. It also infects birds and mammals besides affecting human beings. Though the common symptoms of corona virus are common cold, cough etc., but it is accompanied by severe acute chronic respiratory disease along with multiple organ failure leading to human death. Severe Acute Respiratory Syndrome (SARS) and Middle East Respiratory Syndrome (MERS) are the two major outbreaks in 2003 and 2012 respectively before SARS-CoV2. The source of origin of SARS was located in Southern China. Its fatality rate was within 14%-15%[8] due to which 774 people lost their lives among 8804 affected cases. Saudi Arabia was marked as the base for the commencement of MERS. 858 persons among 2494 infected cases were defeated in their battle against MERS virus. Thus it generated a much higher fatality rate of 34.4%[9] when compared to that of SARS. All of the three epidemic creators SARS, MERS and SARS-CoV2 are biologically included in the genus Betacoronavirus which is under the family Coronaviridae. Both structural and non-structural proteins are involved in the formation of SARS-CoV2. Out of the two, structural proteins like the envelope (E) protein, membrane (M) protein, nucleocapsid (N) protein and the spike (S) protein play a major role in transmitting the disease by binding with the receptors after entering into human body[10]. No clinically proven and approved vaccine for nCoV is available till date though researchers have been trying hard to develop it. So there is an urgent need to understand and analyse the mechanism of disease transmission of this new virus.

Host-pathogen protein-protein interaction networks (PPIN) are very significant for understanding the mechanism of transmission of infection, which is essential for the development of new and more effective therapeutics, including rational drug design. Progression of Infection and disease results in due to the interaction of proteins in between pathogen and host. Pathogen plays an active role in spreading infection. It has been proved to be an acute threat to human lives as it has the mutative capability to adapt itself toward drugs. Pathogen and host PPIN permits pathogenic microorganisms to utilize host capabilities by manipulating the host mechanisms in order to abscond from the immune responses of the host[11–13]. Detection of target proteins through the analysis of pathogen and host PPIN is the central point of research study[14, 15]. Topologically significant proteins having higher degree of interactions are generally found to be essential drug targets. However, proteins having less number of interactions or topologically not significant may be involved in the mechanism of infection because of some biological pathway relevance.

However, clinically validated Human-nCoV protein interaction data is limited in the current literature. This has motivated us to develop a new computational model for nCoV-Human PPI network. We have subsequently validated the proteins involved in the host-pathogen interactions with respect to potential Food and Drug Administration (FDA) drugs for COVID-19 treatment. Key aspects of this research work are highlighted below:

- It has been reported that SARS-CoV has $\sim$ 89%[16, 17] genetic similarities with nCoV.

- SARS-CoV-Human protein-protein interaction network has also been studied widely and available in literature[18–20].

- Recently we have developed a computational model to identify potential spreader proteins in a Human-SARS CoV interaction network using SIS model[21].

- Sequence information of some of the nCoV proteins have been released[22].

- Gene ontological (GO) information (Biological Process (BP), Molecular Function (MF), Cellular Component (CC)) of some of the nCoV proteins are also available[22,23].

- Recently we have also developed a method to predict interaction affinity between proteins from the available GO graph.[24].

- Assessment of interaction affinity between nCoV proteins with potential Human target/bait proteins, which are susceptible to SARS-CoV infection, has been done.

- Fuzzy affinity thresholding is done to detect High Quality nCoV-Human PPIN. The selected human proteins are considered as level-1 human spreader nodes of nCoV.

- Level 2 spreader node in nCoV-Human PPIN are detected using spreadability index and validated by SIS[21, 25] model.

- Validation of our developed model is done with respect to the target proteins of the potential FDA drugs for COVID-19 treatment.[26]

## Results

Our developed computational model of nCoV-Human PPIN contains high quality interactions (HQI) and proteins identified by Fuzzy affinity thresholding and spreadability index validated by SIS model respectively. Sources of input and the generated results always play a crucial role in any computational model which is also true for our proposed model.

**Overview of the data sets**

SARS-CoV-Human PPIN serves as a baseline for our model. The potential level-1 and level-2 human spreaders of SARS-CoV becomes the possible candidate set for selecting level-1 human spreaders of SARS-CoV2. Various datasets have been curated for this purpose which has been outlined below:

*Human PPIN*

The dataset[27, 28] consists of all possible interactions between human proteins that are experimentally documented in humans. Human proteins are represented as nodes while the physical interactions between proteins are represented by edges. It is a collection of 21557 nodes and includes 342353 edges/interactions.

### SARS-CoV PPIN

The dataset[18] consists of interactions between SARS-CoV proteins. It contains 7 unique proteins along with the involvement of 17 interacting edges out of which only the densely connected proteins are considered rather than the isolated ones since theformer play a more active role in transmission of infection than the later.

### SARS-CoV-Human PPIN

The dataset[18] comprises of 118 interactions between SARS-CoV and Human. It is used to fetch the level-1 human interactions of SARS-CoV.

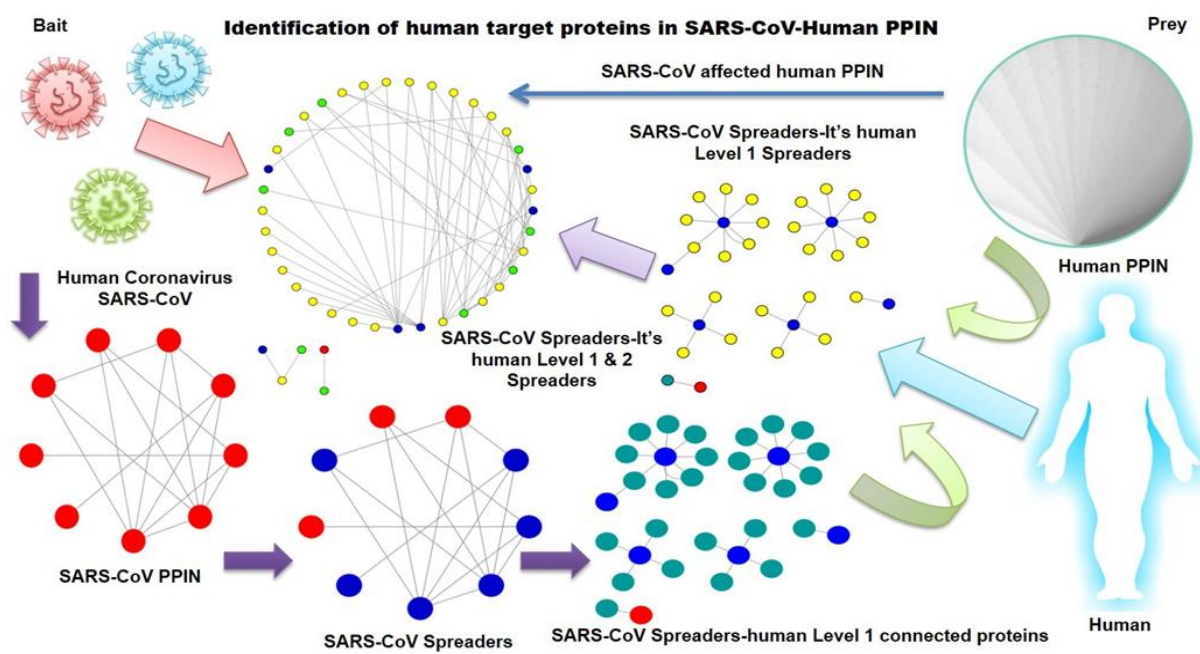

**Figure 1.** Computational model for the selection of spreader nodes in Human-SARS CoV PPIN by spreadability index. Red colored nodes represent SARS-CoV proteins while blue colored nodes are the selected spreader nodes in it. Deep green colored nodes represent level-1 human connected proteins with SARS-CoV proteins while yellow colored nodes represent the selected human spreaders in it. Light green colored nodes represent level-2 human spreaders of SARS-CoV.

### SARS-CoV2 Proteins

This data is collected from pre-released dataset of available SARS-CoV2 protein from UniProtKB[22] (https://covid-19.uniprot.org/) which include 14 reviewed SARS-CoV2 proteins.

### GO Graph and Protein-GO annotations

Three types (CC, MF and BP) of GO graph are collected from GO Consortium[23, 29] (http://geneontology.org/). Protein to GO-annotation map is retrieved from UniProtKB database.

### Potential COVID-19 FDA drugs

Seven potential FDA drugs : Lopinavir[30], Ritonavir[31], Hydroxychloroquine[32, 33], Azithromycin[33], Remdesivir[34–36], Favipi-ravir[37, 38] and Darunavir[39] have been identified from the DrugBank[40] published white paper[26] which have been used forvalidation in our proposed model.

### Selection of spreader nodes in Human-SARS CoV interaction network using spreadability index validated by SIS model:

SARS-CoV-Human PPIN (up to level-2) is formed by the combination of SARS-CoV-Human and Human-Human PPINdatasets. SARS-CoV-Human dataset generates the direct level-1 human interactions of SARS-CoV while human-Human PPIN dataset is used to fetch the corresponding level-2 human interactions. Potential spreader nodes are identified using spreadability index which has been validated by SIS model (see Figure 1)[21]. The selected spreader nodes in SARS-COV-Human PPIN are highlighted in Table S1, Table S2 and Table S3 in the supplementary document. The network view of SARS-CoV-Human PPIN at each level and various selected thresholds are also available online (SARS-CoV Level-1 human spreaders, Level-1 & Level-2:high human spreaders at high threshold of spreadability index, and Level-1 & Level-2:low human spreaders at low threshold of spreadability index).

**Figure 2.** Schematic diagram of Fuzzy PPI model. A) The fuzzy PPI model finds the interaction affinity between the SARS-CoV2 and Human proteins (L1 and L2 spreader of SARS-CoV) using gene ontological information. B) All GO pair-wise interaction affinities are assessed from three independent GO-relationship graphs CC, MF and BP and fuzzy binding affinity of a protein pair is computed from all three pairwise scores of all GO-pair affinities. C) Heatmap representation of Fuzzy PPI score. D) Network representation of Human and SARS-CoV2 proteins with 0.2 onward threshold of Fuzzy PPI score at high specificity. Finally, high quality interactions are extracted to retrieve the potential human prey for SARS-CoV2 at 1.4 threshold.

**Figure 3.** Network representation of HQIs (score≥ 0.4) between SARS-CoV2 and human proteins. Blue and yellow spherical nodes represent the SARS-CoV2 and human proteins respectively. The edge width reflects the fuzzy score of binding affinity.

**Identification of the nCoV-Human proteins interactions using Fuzzy PPI model:**

The GO information can be useful to infer the binding affinity of any pair of interacting proteins using three different types of GO hierarchical relationship graph (CC, MF and BP)[23]. The fuzzy PPI model has been proposed to find the interaction affinity between the SARS-CoV2 and Human proteins using GO based information (please see Figure 2 and Methods for details). To identify the interactors of SARS-CoV2 on human using the Fuzzy PPI model, a set of candidate proteins are selected as the L1 and L2 spreader nodes of SARS-CoV using SIS model (as depicted in Figure 1). Fuzzy PPI model is constructed from the ontological relationship graphs by evaluating the affinity between all possible GO pairs annotated from any target protein pair and finally the fuzzy score of interaction affinity of protein pair is computed from these GO pair-wise interaction affinity in to a range of [0,1] (details are discussed in the Methods). The heatmap representation of fuzzy interaction affinities (with score ≥ 0.2 for very high specificity ~ 99%) is shown in supplementary Figure S1 and Table S4. The high quality interaction (HQI) is retrieved at threshold 0.4, (almost ~ 99.98% Specificity) which results total 78 interactions between SARS-CoV2 and Human (37 proteins). The interaction networks predicted from Fuzzy-PPI model are shown in Figure 3.

**Identification of Human spreader proteins for nCoV**

Human proteins present in the high quality interactions of nCoV-Human PPIN fetched by applying fuzzy affinity threshold are considered as level-1 spreaders. From these level-1 spreaders, corresponding level-2 human interactions are obtained using human-Human PPIN dataset. Spreadability index is thus computed for these level-2 human proteins for the identification of level-2 human spreader nodes. The selection is also verified by SIS model. The selected spreader nodes in SARSCOV-Human PPIN are highlighted in Table S4, Table S5 and Table S6 in the supplementary document. A sample computational model of nCoV-Human PPIN under high threshold has been highlighted in Figure 4. The network view of SARS-CoV2-HumanPPIN at each level and various selected thresholds are also available online (SARS-CoV2 Level-1 human spreaders, Level-1 & Level-2:high spreaders at high threshold of spreadability index and Level-1 & Level-2:low human spreaders at low threshold of spreadability index).

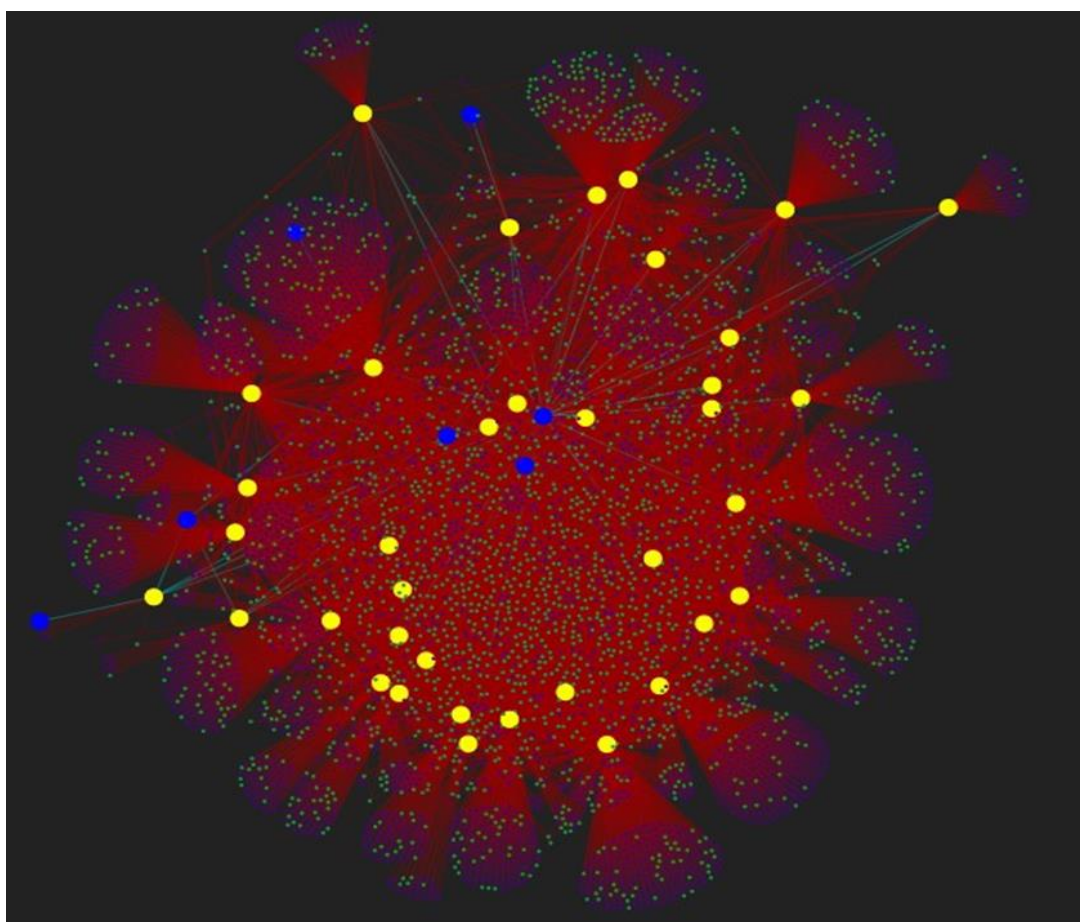

**Figure 4.** Derived nCoV-Human PPIN with human spreader proteins from proposed computational model. Blue, yellow and green colored nodes denote nCoV spreaders, its human level-1 and level-2 spreaders. level-1 human spreaders are detected by applying fuzzy affinity thresholding while level-2 human spreaders are identified by spreadability index validated by SIS model.

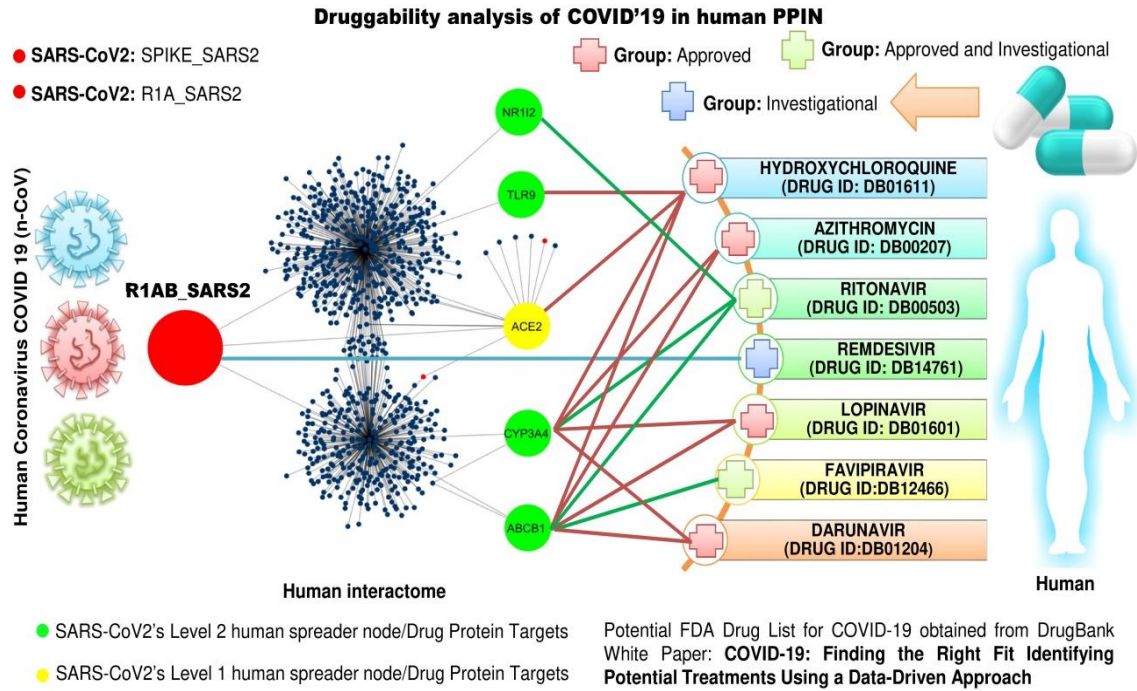

**Figure 5.** Validation of our developed computational model with respect to the target proteins of the FDA accepted drugs for COVID-19 treatment. Yellow and green colored nodes denote level-1 and level-2 human spreader of nCoV which acts as the drug protein targets.

### Validation using potential FDA drugs for COVID-19:

After proper assessment of all potential drugs as mentioned in the DrugBank[40] white paper[26], seven drugs Lopinavir[30], Ritonavir[31], Hydroxychloroquine[32, 33], Azithromycin[33], Remdesivir[34–36], Favipiravir[37, 38] and Darunavir[39] are identified which are showing expected results to some extent in the clinical trials done for SARS-CoV2 vaccine till date. All approved human protein targets for each of the five approved drugs are fetched from the advanced search section[41] of drug bank[40, 42]. These targets when searched in our proposed model of nCoV-Human PPIN are found to play an active role of spreader nodes. This reveals the fact that the selected spreader nodes are of biological importance in transmitting infection in a network which actually make them the protein drug targets of the potential FDA drugs for COVID-19. The target protein hits in our nCoV-Human PPIN for each of the 7 potential FDA drugs are highlighted in Figure 5. It can be observed that 4 target protein hits are obtained for Hydroxychloroquine, 3 target proteins for Ritonavir, 2 target protein hits for each of Lopinavir, Darunavir and Azithromycin and 1 target protein hit for Remdesivir and Favipiravir. Out of these protein targets, ACE2 is the most important one since it is considered to be the one of the crucial receptors of human for nCoV to transmit infection deep inside the human cell[43–45].

## Discussion

In any host-pathogen interaction network, identification of spreader nodes is crucial for disease prognosis. Not every protein in an interaction network has intense disease spreading capability. In this work, we have used the SARS-CoV-Human PPIN network and the spreader nodes at both level-1 and level-2 using the SIS model. These spreader nodes are considered for computing the protein interaction affinity score to unmask the level-1 human spreaders of nCoV. GO annotations have been also taken into consideration along with PPIN properties to make this model more effective and significant. With the gradual progress of the work, it has been observed that the selected human spreader nodes, identified by our proposed model, emerge as the potential protein targets of the FDA approved drugs for COVID-19. The basic hypotheses of the work may be listed as follows:

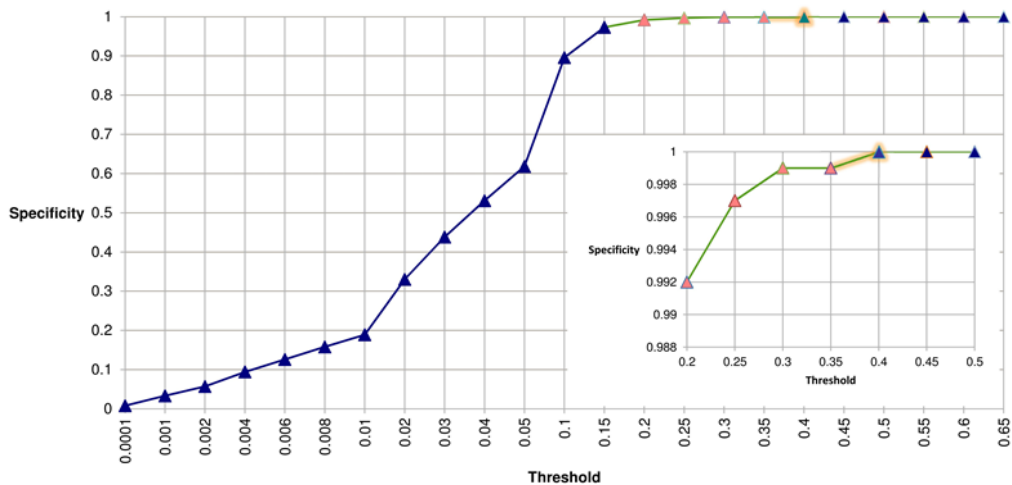

**Figure 6.** Specificity at different threshold (x-axis) of binding affinity obtained from Fuzzy PPI model for complete human proteome interaction network. At 0.2 onward threshold, it produces high specificity with respect to benchmark positive and negative interaction data. High quality interactions are extracted at 0.4 threshold with ~ 99.9% specificity.

- There is a genetic overlap of ~ 89% (as suggested by ICTV) between SARS-CoV and SARS-CoV2, which also leads to a significant overlap in spreader proteins between human-SARS-COV and human-SARS-COV2 protein-interaction network.
- Fuzzy PPI approach can assess protein interaction affinities at very high specificity with respect to benchmark datasets
  as shown in Figure 6. High specificity signifies very low false positive rate at a given threshold. Thus, at 0.4 threshold (~ 99.9% specificity), the proposed model evaluates high quality positive interactions in Human-nCoV PPIN.

Finally we propose, that the developed computational model effectively identifies Human-nCoV PPIs with high specificity. nCoV-Human interactions are inferred from another pandemic initiator SARS-CoV which is believed to be highly genetically similar to nCoV. We also identify the human spreader proteins (up to level-2) using spreadability index, validated through SIS model. Due to high network density in human interaction network, number of proteins increase with the transition from one level to another. So, our proposed model is also capable of identifying human spreader proteins in level-2 by using spreadability index which is validated by SIS model.

Target proteins of the potential FDA drugs for COVID-19 are found to overlap with the spreader nodes of the proposed computational nCoV-Human protein interaction model. Target proteins of seven potential FDA drugs: Lopinavir[30], Ritonavir[31], Hydroxychloroquine[32, 33], Azithromycin[33], Remdesivir[34–36], Favipiravir[37, 38] and Darunavir[39] for COVID-19 as mentioned in the DrugBank white paper[26] overlap with the spreader nodes of the proposed in silico nCoV-Human protein interaction model (see Figure 5). Though clinical trials for COVID-19 vaccine in 2020 are on their way till date but three out of the seven *i.e.* Remdesivir[46] Hydroxychloroquine[47] and Favipiravir[47] are found to be the most promising as well as effective ones and their protein targets R1AB_SARS2, TLR9, ACE2, CYP3A4 and ABCB1 are also successfully identified as spreader nodes by our proposed model. This assessment reveals the fact that these spreader nodes are indeed have biological relevance relative to disease propagation.

## Methods

Our developed computational model for nCoV-Human PPIN consists of two important methodologies 1) identification of spreader nodes by spreadability index along with the validation of SIS model and 2) Fuzzy PPI model.

**Identification of spreader nodes by spreadability index along with the validation of SIS model:**
In nCoV-Human PPIN, the former acts as a pathogen while the host acts as bait. The transmission of infection starts when a pathogen enters a host body and starts infecting its protein which in turn affects its directly or indirectly connected neighborhood proteins. Considering this method of transmission, PPIN of human and SARS-CoV are considered to detect spreader nodes. Spreader nodes are those nodes proteins which actually transmits the disease fast among its neighbors. But not all the nodes in a PPIN are spreaders. So, proper detection of spreader nodes is crucial. Thus, spreader nodes are identified by spreadability index which actually measure the transmission capability of a node protein.

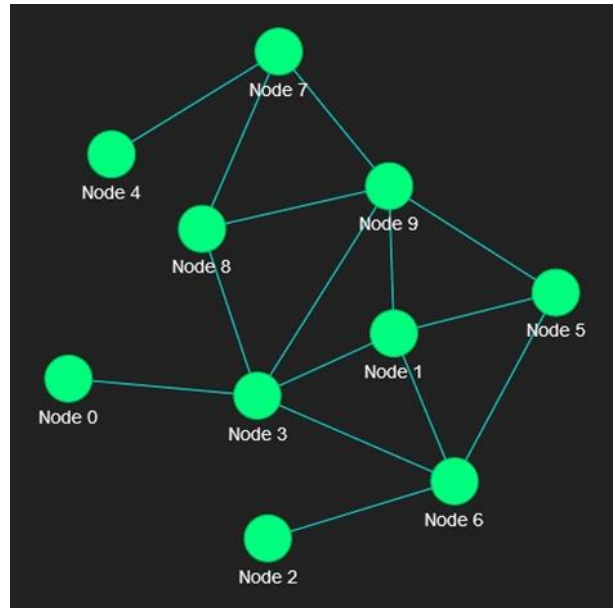

**Figure 7.** Synthetic protein-protein interaction network. It is comprised of 10 nodes and 25 edges.

**Table 1.** Computation of spreadability index of Figure 7 along with validation of selected top 5 spreader nodes by SIS model.

| Rank | Proteins | $E_{out}^{S_i}$ | $E_{in}^{S_i}$ | Edge Ratio | Neighborhood Density | Node Weight | Spreadability Index | Sum of SIS infection rate of top 5 nodes |
|---|---|---|---|---|---|---|---|---|
| 1 | Node 3 | 6 | 3 | 1.75 | 6.94 | 2.83 | 14.99 | **1.19** |
| 2 | Node 9 | 5 | 4 | 1.20 | 7.07 | 3.00 | 11.48 | |
| 3 | Node 6 | 5 | 2 | 2.00 | 3.93 | 2.60 | 10.46 | |
| 4 | Node 8 | 6 | 2 | 2.33 | 2.27 | 3.25 | 8.55 | |
| 5 | Node 1 | 5 | 4 | 1.20 | 4.21 | 3.40 | 8.45 | |

Compactness of PPIN and its transferal capability is evaluated using centrality analysis. Nodes having high centrality value are usually considered as spreader nodes or the most critical node in a network.

Spreadability index[21] is one of the centrality based measure which is a combination of three major topological neighborhood based features of a network. They are: 1) Node weight[48] 2) Edge ratio[49] and 3) Neighborhood density[49]. Nodes having high spreadability index are considered as spreader nodes. The spreader nodes thus identified are also validated by SIS model[25]. SIS model is implemented with a motive of designing the SARS-CoV and SARS-CoV2 outbreak into a disease model consisting of proteins based on their present infection status. A protein can be in either of one of the three states: 1) S: Susceptible, which means that every protein is initially susceptible though it is not yet infected but is at risk of getting infected by disease. 2) I: Infected, which means that the protein is already infected by the disease and 3) S: Susceptible, which means proteins again become susceptible after getting recovered from infected state. This model is implemented in such a way that it generates the overall infection capability of a node after a certain range of iterations. Thus the sum of the infection capability of the top selected spreader nodes are computed by this model which is compared against the sum obtained for the selected top critical nodes by other existing centrality measures like Betweenness Centrality (BC)[50], Closeness centrality (CC)[51], Degree centrality (DC)[52] and Local average centrality (LAC)[53]. Our proposed method of selecting spreader nodes[21] has performed better in comparison to the other existing

*state-of-the-art*. A synthetic PPIN is considered in Figure 7 to demonstrate the entire methodology of spreadability index. Computational analysis of spreadability index of our proposed model with one of the other methodology BC has been highlighted from Table 1 to supplementary Table S7. $E_{out}^{S_i}$ is the total number of edges which are outgoing from the ego network $S_i$ whereas $E_{in}^{S_i}$ is denoted as the total number of interconnections in the neighborhood of node $i^{49}$. $E_{out}^{S_3}$ of node 3 is 6 while $E_{in}^{S_3}$ of node 3 is 3 which highlights the fact that node 3 has the highest transmission ability from its ego network to outside when compared to other nodes. Node 3 has also the highest spreadability index. But BC failed to rank node 3 in first position. Same scenario can be observed for some other nodes in the synthetic network too. Besides SIS validation result shows that the selected top ranked spreader nodes in this proposed model have the highest infection capability in comparison to the other ranked nodes.

**Fuzzy PPI Model for potential SARS-CoV2-Human interaction identification:**
The binding affinity between any two interacting proteins can be estimated by combining the semantic similarity scores of the GO terms associated with the proteins[15, 24, 54–56]. Greater number of semantically similar GO annotations between any protein pair indicates higher interaction affinity. Fuzzy PPI model is a hybrid approach[24] that utilizes both the topological[57] features of the GO graph and information contents[56, 58, 59] of the GO terms.

GO is organized in three independent directed acyclic graphs (DAGs): molecular function (MF), biological process(BP), and cellular component(CC)[23]. The nodes in each GO graph represent GO terms and the edges represent different hierarchical relationships. In this work, two most important relations *'is a'* and *'part of'* has been used for GO relations[29].

The semantic similarity between any two proteins is estimated by considering the similarities between their all pairs of annotating gene ontology (GO) terms belonging to a particular ontological graph. The similarity of a GO term pair is determined by considering certain topological properties (shortest path length) of the GO graph and the average information content (IC)[60] of the disjunctive common ancestors (DCAs)[54, 55] of the GO terms as proposed in[24]. Fuzzy PPI first rely on a fuzzy clustering of the GO graph where the selection of GO terms as cluster center is based on the level of association of that GO term in the GO graph. The cluster centers are selected based on the proportion measure of GO terms. The proportion measure for any GO term t is computed as $PrM(t) = (|An(t)| + |Dn(t)|)/|O(t)|$ where $An(t), Dn(t)$ represents the ascendant and descendant of term $t$ and $O(t)$ is the total number of GO terms in ontology O. Higher value of proportion measure signifies higher coverage of ascendants and descendants associated with the specific node. The GO terms for which this proportion measure is above a predefined threshold are selected as cluster centers. In this work, the cluster centers are selected based on the threshold values as suggested in [15, 24].

After selecting the cluster centers, the degree of membership of a GO term to each of the selected cluster centers is calculated using its respective shortest path lengths to the corresponding cluster centers. The membership of the GO term to a cluster decreases with increase in its shortest path length to the cluster center. The membership function is defined as $MmF_c(t) = e^{-(x-c_i)^2/2k^2}$, where $c_i$ is $i-th$ center and $k$ is the width of membership function and $x$ is the shortest path length from $t$ to $c_i$. The difference in membership values between the GO pair $t_i$ and $t_j$ with respect to each cluster center, are computed to find the weight parameter. The weight parameter is defined as $Wt(t_i, t_j) = 1 - maxD(t_i, t_j)$. This weight value determines how dissimilar two GO terms can be with respect to the cluster centers. Next, the shared information content (SIC) is computed using average IC[60], of the DCAs of GO term pair $(t_i, t_j)$ is determined from three GO graphs. The SIC is defined as $SIC(t_i, t_j) = \sum_{a \in DCA(t_i, t_j)} IC(a)/|DCA(t_i, t_j)|$, where $DCA(t_i, t_j)$ represents the disjunctive common ancestors of GO-term $t_i$ and $t_j$. The semantic similarity of protein pair $(P_i, P_j)$ for each GO-type (CC, MF and BP), is estimated by utilizing the maximum similarity of all possible GO pairs from the annotations of proteins $P_i$ and $P_j$ for each type of GO. The binding affinity of protein pair $(P_i, P_j)$ is defined as the average of CC, MF and BP based semantic similarity. The fuzzy score of binding affinity is computed by normalizing the binding affinity using max-min normalization. In this work, the fuzzy binding affinity score is computed between the protein pairs of SARS-CoV2 and human proteins using the available ontological information. Finally, with high specificity threshold (please see Figure 6), high quality interactions are extracted for human-SARS-CoV2.

## Acknowledgements

This work is partially supported by the CMATER research laboratory of the Computer Science and Engineering Department, Jadavpur University, India, PURSE-II and UPE-II grants. Subhadip Basu acknowledges Department of Biotechnology grant (BT/PR16356/BID/7/596/2016), Government of India. For research fellowship support, Anup Kumar Halder acknowledges the Visvesvaraya PhD Scheme for Electronics & IT, an initiative of Ministry of Electronics & Information Technology (MeitY), Government of India. We acknowledge Prof. Jacek Sroka (Institute of Informatics, University of Warsaw ) for his contribution toward the developments of the fuzzy ppi methods.

## Author contributions statement

S.S., A.K.H. and S.B. conceived the idea of the research and wrote the manuscript. S.S. and A.K.H. conducted the experiment(s). P.C, S.S.B., M.N. and S.B. analyzed the results. M.N., P.C and S.B. reviewed the manuscript.

## Additional information

**Competing interests:** The authors declare no competing interests.

All [Supplementary information](#) are freely available for academic and research purpose only.

All queries should be send to the corresponding author's email: subhadip.basu@jadavpuruniversity.in